# Characterizing Exoplanets for Assessing Their Potential Habitability

J. M. Rodríguez-Mozos and A. Moya
Departament d'Astronomia i Astrofísica, Universitat de València, C. Dr. Moliner 50, 46100 Burjassot, Spain; josemaria.rodriguezmozos@gmail.com, andres.moya-bedon@uv.es

## Abstract

The Statistical-likelihood Exoplanetary Habitability Index (SEPHI) serves as a valuable tool for prioritizing targets for further study and identifying potentially habitable environments. In this paper, we present SEPHI 2.0, which incorporates several key improvements: (1) updated methods for estimating exoplanet internal structures and magnetic fields; (2) the inclusion of orbital eccentricity in assessing the potential for liquid water on an exoplanet's surface; and (3) a new exoplanet mass–radius relationship. SEPHI 2.0 retains its probabilistic framework and combines the different subindexes by selecting the most restrictive one. In SEPHI 2.0, atmospheric retention is consolidated into a single index that incorporates both thermal (Jeans escape) and nonthermal (stellar wind and magnetic effect) processes. Recent advancements in estimating exoplanet internal structures and magnetic fields have been integrated. Additionally, a new empirical exoplanet mass–radius relationship is introduced. All this is incorporated into the SADE code, which uses key data on exoplanets and their host stars to assess habitability and prioritize targets for further study, providing a comprehensive output of an exoplanetary system's physical characteristics. SADE is available as a free online tool. Notably, this is the first approach to including estimated exoplanet magnetic fields in a habitability index. The SADE software facilitates the identification of potentially habitable exoplanets. Among the 5500+ confirmed exoplanets, only a few—such as Kepler-62f and GJ 514b—achieve SEPHI 2.0 scores close to 1. It is also noticeable that, according to our studies, TRAPPIST-1 f and g are ranked higher than TRAPPIST-1 e in terms of habitability potential.

*Unified Astronomy Thesaurus concepts:* Exoplanet atmospheres (487); Exoplanet structure (495); Astrobiology (74); Exoplanet catalogs (488)

*Materials only available in the* online version of record: *machine-readable table*

## 1. Introduction

The search for life beyond the boundaries of our solar system stands as one of the most exciting scientific challenges we have faced since a few decades ago. Advancements in observational techniques and the discovery of thousands of exoplanets have transformed this endeavor into a rigorous scientific pursuit.

Exoplanets—planets circling stars other than the Sun—reveal the remarkable variety of planetary systems in our galaxy. Some are hot gas giants orbiting extremely close to their stars; others are cold, distant ice worlds. This breadth of environments challenges current ideas about how planets form and evolve. Among these diverse worlds, we may find conditions that foster the emergence and sustenance of life.

In the pursuit of understanding the prevalence and potential habitability of exoplanets, the concept of the habitability index has emerged as a crucial metric for assessing the suitability of distant worlds to support life as we know it. As soon as the first small planets close to their stellar habitable zone (HZ) were discovered, the necessity of an index comparing them arose (D. Schulze-Makuch et al. 2011). In the abovementioned work, the authors presented two indexes: (1) the Earth Similarity Index (ESI; Equation (1)), which physically compares an exoplanet with the Earth; and (2) the Planetary Habitability Index (PHI; Equation (2)), which accounts for the habitability potential of an exoplanet compared with what we know about life. Therefore, the goal of these indexes is to define a classification scheme for ascertaining whether exoplanets are potentially habitable from an astrobiological standpoint.

In J. M. Rodríguez-Mozos & A. Moya (2017), we proposed a new strategy for assessing the potential habitability of exoplanets. This Statistical-likelihood Exoplanetary Habitability Index (SEPHI) has been instrumental in quantifying the resemblance of exoplanets, offering a valuable tool for prioritizing targets for further study and potentially identifying environments conducive to life beyond our solar system. The search for habitable exoplanets has witnessed remarkable progress in recent years, helped by advancements in observational techniques and data analysis methods. With thousands of exoplanets detected to date, ranging from gas giants to rocky terrestrial worlds, the focus has shifted toward characterizing the properties of these distant worlds and assessing their potential habitability. In this field, SEPHI has emerged as a useful metric for evaluating the habitability of exoplanets, in terms of the type of exoplanet, composition, and orbital characteristics. SEPHI enables astronomers to prioritize targets for follow-up observations and identify promising candidates for further study.

In this paper, we present a new version of SEPHI (SEPHI 2.0), in which a remarkable number of improvements have been made. We incorporate new developments related to the estimation of the internal structures of exoplanets and their magnetic fields, presented in J. M. Rodríguez-Mozos & A. Moya (2019) and J. M. Rodríguez-Mozos & A. Moya (2022). In addition, a new definition of some subindexes and a new combination of them is proposed. The inclusion of the orbital eccentricity in the definition of the presence of liquid water on the exoplanet's surface and a new exoplanetary mass–radius relation complete the list of improvements and new developments presented in this work.

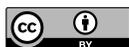

## 2. A New Version of SEPHI: SEPHI 2.0

J. M. Rodríguez-Mozos & A. Moya (2017) proposed a new method for assessing planetary habitability, with the following features:

1. SEPHI should have a probabilistic interpretation, allowing for easier and more accurate result interpretation.
2. The index compares four characteristics of exoplanets that describe the basic conditions that can allow life: being a rocky planet, having the capacity to retain an atmosphere, supporting liquid water on the surface, and having a magnetic field to shield against stellar winds.
3. Each characteristic is represented by individual probability functions. SEPHI is calculated as the geometric mean of the likelihood of these characteristics.

In the studies by J. M. Rodríguez-Mozos & A. Moya (2019, 2022), the authors address the accurate estimation of the impact of stellar activity on exoplanet habitability. These studies focus on common exoplanet and stellar characteristics, such as masses, radii, stellar effective temperatures, and orbital periods, to name a few. The first study estimates the internal structures of known exoplanets and the effects of stellar winds on them, assuming the exoplanet magnetic field. This led to the systematic estimation, for the first time, of the apertures of the auroral rings of exoplanets. They found that a stronger exoplanetary magnetic field reduces the area of the unprotected atmosphere, leading to a smaller auroral angle. Additionally, stronger dynamic and magnetic pressures from the star increase the unprotected surface area, resulting in greater atmospheric erosion. The second study addresses the estimation of exoplanetary magnetic fields, offering, also for the first time, a systematic and massive estimation of exoplanetary magnetic fields if the planetary dynamo is active. These two studies allowed for the probability of an exoplanet maintaining its atmosphere as a balance between its magnetic field and stellar pressure.

This has led to a new version of SEPHI (SEPHI 2.0), which retains its probabilistic interpretation and includes the following improvements:

1. The way the final probability is obtained from the different criteria likelihoods (Sections 2.1 and 2.5).
2. An update of the thermal escape mechanism has been included (Section 2.3.1).
3. A joint coefficient, where thermal and nonthermal space mechanisms are considered together (Section 2.3.3).
4. The orbital eccentricity is now included when analyzing the stellar HZ (Section 2.4).

As a consequence, SEPHI 2.0 now focuses on three main criteria for potential habitability:

1. Being a rocky planet.
2. Retaining an atmosphere.
3. Supporting liquid water on the surface.

### 2.1. New Combination of Subindexes

One common approach to combining subindexes, in our case likelihoods, is through geometric averaging, as done by SEPHI and other indices, like the PHI (D. Schulze-Makuch et al. 2011). However, after reviewing numerous exoplanets from the literature, we found that geometric averaging is not suitable for estimating a planet's potential habitability.

To assess habitability, SEPHI 2.0 uses three criteria. For a planet to be potentially habitable, it must simultaneously meet all three conditions: being a rocky planet, retaining an atmosphere, and supporting liquid water. A geometric mean can mask deficiencies in one criterion by a high likelihood in another. For instance, a planet with a low probability of retaining an atmosphere but a high likelihood of having liquid water may still yield a moderate habitability score, whereas it should be low, due to the critical absence of an atmosphere.

Therefore, in SEPHI 2.0, the new assumption is that the main driver for assessing potential habitability is the most restrictive criterion. That is, we define the final index value as

$$\text{SEPHI 2.0} = \min(L_1, L_2, L_3), \quad (1)$$

where $L_1$, $L_2$, and $L_3$ represent the likelihood functions for being a rocky planet, retaining an atmosphere, and having liquid water, respectively. To identify the best candidates for hosting life, we should seek exoplanets with SEPHI 2.0 values closest to unity.

In the following sections, we will describe the three criteria and the improvements included in SEPHI 2.0 with respect SEPHI.

### 2.2. $L_1$, Being a Rocky Planet

We will typically work with four types of planets: dry rocky planets, water-rich rocky planets, ice giants, and gas giants. To classify these planets, we use a grid of models based on their mass and radius, which helps determine the planet type and its estimated composition. The initial version of SEPHI utilized the grid proposed by L. Zeng & D. Sasselov (2013). However, since this grid does not accurately reflect Earth's internal composition, we have developed a grid based on the Earth's internal structure, as described by A. M. Dziewonski & D. L. Anderson (1981), for dry rocky exoplanets. Additionally, a grid for water-rich exoplanets has been developed using observational data from Ganymede. This grid is detailed in Appendix B of J. M. Rodríguez-Mozos & A. Moya (2022).

Before any numerical assessment, it is important to discuss the potential habitability of water-rich rocky planets from the point of view of the only successful case we know, the Earth. In our solar system, life emerged on only one dry rocky planet and possibly on Mars, before it lost its magnetic field. Since there are no water-rich rocky planets in our solar system in the HZ, we cannot directly compare the development of life on Earth with a hypothetical water-rich rocky planet in the HZ.

Moreover, if one of Jupiter's moons, such as Europa or Ganymede (both water-rich rocky bodies), were to enter the Sun's HZ due to gravitational interactions and assume an orbit similar to Earth's, they would receive a similar energy flux as Earth does from the Sun, the only solid zone on the planet surface would be polar ice caps, more or less wide, with the rest of the surface being liquid water. In addition, certain ocean planets, like Kepler-22 b, located in the HZ of their star, might produce high amounts of water vapor in their atmospheres. This could hinder heat dissipation from the planet's surface, changing the physical conditions on its surface. The absence of a solid dry surface and these different physical conditions would make life potentially very different from Earth. For this reason, water is essential for life; even a small amount of water in the composition of a planet can help in its development, but high amounts of water do not seem to contribute to greater habitability concerning dry rocky planets. This baseline applies to SEPHI 2.0. If future research clarifies the potential habitability of water-rich planets, we can easily adjust the index to incorporate that new information.

To determine the probability function, we use two properties of Gaussian distributions. First, the maximum probability occurs when the variable equals the distribution's mean ($\mu$), which will be 1 in this case. Second, the probability is approximately 0 when the variable equals $\mu \pm 3\sigma$.

Suppose a water-rich planet exists with a mass of $8.2\,M_\oplus$ (see Section 4) and an ice mass fraction of 0.2; according to our mass–radius grid (J. M. Rodríguez-Mozos & A. Moya 2022), the outer layer of ice and water could reach 2300 km. If we admit that the origin of life on Earth occurred in hydrothermal fumaroles located between the Earth's crust and the ocean (M. Dodd et al. 2017; V. Helmbrecht et al. 2017), at that depth of 2300 km, pressure and temperature values could complicate the conditions for the beginning of life. For this model, a dry rocky planet or a water-rich rocky planet with up to 20% of its mass in water will have a probability of 1 for supporting life, meaning if its radius is less than or equal to the radius corresponding to 20% water mass ($R_{20}$). Therefore, we set $\mu_1 = R_{20}$. To improve the habitability of water-rich planets compared to mini-Neptunes, exoplanets with a composition of 100% water have been assigned a minimum probability of 0.2. Thus, defining $L_1(R_{100}) = 0.2$, we can easily determine that

$$\sigma_1 = \frac{R_{100} - R_{20}}{\sqrt{-2\ln(0.2)}}. \qquad (2)$$

With this, the probability function will have the form

$$L_1(R_p) = \begin{cases} 1 & \text{for } R_p \leqslant R_{20} \\ e^{-\frac{1}{2}\left(\frac{R_p - \mu_1}{\sigma_1}\right)^2} & \text{for } R_p > R_{20} \end{cases}. \qquad (3)$$

### 2.3. $L_2$, Retaining an Atmosphere

In the initial version of SEPHI, the probability of a planet retaining its atmosphere was assessed solely based on the planet's escape velocity, without considering that higher energy levels in atmospheric molecules can facilitate atmospheric escape mechanisms. Nevertheless, there are multiple atmospheric escape mechanisms. They can be classified into thermal and nonthermal escape mechanisms. In this work, we study the two predominant mechanisms on Earth at this time, which are Jeans escapement (the thermal mechanism) and stellar wind (the nonthermal mechanism).

#### 2.3.1. $L_{21}$, Jeans Escape

The average velocity of gas atoms or molecules depends on their temperature and mass. When atmospheric gas components reach the exosphere, collisions between particles become negligible, allowing those with higher velocities to overcome gravitational forces and escape into space. Lighter atmospheric components like hydrogen and helium are more likely to escape through this mechanism. This is the so-called Jeans escape.

It is important to note that this escape occurs atom by atom and molecule by molecule, typically taking billions of years. However, other more drastic thermal escape processes, such as hydrodynamic escape, can occur, especially around young, rapidly rotating stars emitting significant ultraviolet radiation. This mechanism can cause the mass escape of certain atmospheric components over tens of millions of years and can be particularly intense around active M-type stars that emit X-rays and extreme-ultraviolet radiation for extended periods. This radiation leads to the heating and expansion of the upper atmospheric layers of planets in close orbits (H. Lammer et al. 2007).

To quantify Jeans escape, we first determine the equilibrium temperature of the planet ($T_{\rm eq}$). For some exoplanets, $T_{\rm eq}$ is available in the NASA Exoplanets Archive (NEA; NASA Exoplanet Archive 2019). When $T_{\rm eq}$ is unknown, it can be calculated using the stellar data and the planet's semimajor axis with the following equation:

$$T_{\rm eq} = T_{\rm eff}(1 - A)^{1/4}\sqrt{\frac{R_s}{2a}}, \qquad (4)$$

where $T_{\rm eff}$ is the effective temperature of the star, $A$ is the planet's albedo, $R_s$ is the star's radius, and $a$ is the semimajor axis of the planet's orbit.

The above equation indicates that $T_{\rm eq}$ is not very sensitive to albedo variation. For example, if we modify the Earth's albedo by $\pm 30\%$, the $T_{\rm eq}$ obtained for the Earth would only be modified by $\pm 3\%$. For this reason, generic values of the albedo have been used according to the type of planet. For dry rocky planets, the Earth's albedo ($A = 0.3$) has been used, while for water-rich planets and giant planets, Jupiter's albedo ($A = 0.41$) has been used.

Once the equilibrium temperature is determined, we use it to calculate the escape velocity for common atmospheric components like $H_2$, $He$, $CH_4$, $NH_3$, $H_2O$, $N_2$, $O_2$, $CO_2$, and $Xe$, based on data from J. W. Chamberlain & D. M. Hunten (1987). This allows us to determine the escape velocity for each atmospheric component.

Next, we calculate the planet's surface escape velocity ($v_{\rm esc}$), using the formula

$$v_{\rm esc} = \sqrt{\frac{2GM_p}{R_p}}, \qquad (5)$$

where $G$ is the universal gravitational constant, $M_p$ is the planet's mass in kilograms, and $R_p$ is the planet's radius in meters. By comparing the velocities of different atmospheric components with the planet's escape velocity, we can determine which components are likely to escape.

To assess the likelihood of a planet retaining its atmosphere, we consider that if the planet's escape velocity is higher than the escape velocity of helium ($v_{\rm esc}({\rm He})$), no atmospheric components except hydrogen can escape. Hydrogen, being very light, cannot be retained by some rocky planets like Earth. Therefore, we assume that hydrogen escape does not significantly affect life development over long timescales. Therefore, we take $\mu_{21} = v_{\rm esc}({\rm He})$. On the other hand, if a planet's escape velocity is so low that it allows the escape of a heavy molecule like $CO_2$, it implies that lighter molecules like hydrogen, water, and oxygen have already escaped, severely compromising the planet's habitability. Therefore, the probability of retaining the atmosphere is considered to be 0 in this case. Hence,

$$\sigma_{21} = \frac{v_{\text{esc}}(\text{He}) - v_{\text{esc}}(\text{CO}_2)}{3}. \tag{6}$$

Thus, the probability function is defined as follows:

$$L_{21}(v_{\text{esc}}) = \begin{cases} 1 & \text{for } v_{\text{esc}} \geqslant v_{\text{esc}}(\text{He}) \\ e^{-\frac{1}{2}\left(\frac{v_{\text{esc}} - \mu_{21}}{\sigma_{21}}\right)^2} & \text{for } v_{\text{esc}} < v_{\text{esc}}(\text{He}) \end{cases}. \tag{7}$$

*2.3.2. $L_{22}$, Escape via Stellar Winds*

In the context of nonthermal escape, we will analyze the atmospheric erosion caused by stellar winds on planetary atmospheres. Stars with convective outer layers—up to stellar masses between 1.3 and 1.4 $M_\odot$—experience mass loss in the form of stellar winds throughout their lifetimes (A. A. Vidotto 2018). Stellar winds can significantly erode planetary atmospheres, especially for planets in the HZ of M dwarfs, due to their proximity to the star. However, planetary magnetic fields protect atmospheres from stellar wind erosion. The protected zone is defined by the radius of the magnetosphere ($r_M$), and it is determined by the balance between the magnetic pressure from the star and the dynamic pressure of the stellar wind, as well as the magnetic pressure generated by the planet's dynamo. This study follows the methodology of J. M. Rodríguez-Mozos & A. Moya (2019).

An active planetary dynamo generates a magnetic field with closed lines trapping ions. However, near the magnetic poles, these lines open to interplanetary space, allowing ion escape. In addition, these regions are exposed to both the solar wind and cosmic radiation, leading to phenomena such as polar auroras.

The unprotected zone of a planet can be determined by the following equation, as described by (A. A. Vidotto et al. 2013):

$$\alpha_o = \arcsin\left[\left(\frac{R_p}{r_M}\right)^{1/2}\right], \tag{8}$$

with $\alpha_o$ being the opening angle of the auroral ring. The unprotected zone is part of the atmosphere delimited by a cone of revolution generated by a line that, passing through the center of the planet, forms an angle $\alpha_o$ with the magnetic axis of the planet. For the current conditions of the Sun, the magnetosphere radius on Earth's dayside is approximately $10\,R_\oplus$, and the unprotected region on Earth can be estimated with an opening angle of $\alpha_o = 18°$ from the magnetic poles.

As the magnetic pressure from the star or the dynamic pressure of the stellar wind increases, the magnetosphere's size decreases, expanding the unprotected area, as indicated by Equation (8). This reduction in magnetosphere size correlates with an increase in atmospheric escape from the exosphere, potentially altering the atmospheric composition over time.

The oldest records of the Earth's magnetic field, based on a thermoremanent magnetization of silicate crystals, indicate that the Earth's geodynamo was active at least 3.4 Ga ago, with the magnetic field strength being measured between 50% and 70% of today's field (J. A. Tarduno et al. 2010). Stellar models predict a much faster rotation period for the Sun at that time, with greater X-ray emission and greater mass loss than current ones, which eventually produced greater pressure from the solar wind on Earth's magnetopause (J. A. Tarduno et al. 2010). In those conditions—that is, 3.4 Ga ago and with the planetary dynamo already active—the size of the magnetosphere came to be around $5\,R_\oplus$, or even slightly below, for a long time (J. A. Tarduno et al. 2010). Even in this case, the early Earth retained its water and atmosphere and thus evolved as a habitable planet (J. A. Tarduno et al. 2010). For this reason, it will be considered that any planet with a magnetopause radius greater than this value will be able to maintain its atmosphere without problems—that is,

$$\mu_{22} = r_{\text{M early Earth}} \approx 5R_p. \tag{9}$$

When the magnetopause radius equals the planet's radius, the magnetic protection is lost, exposing the entire atmosphere to stellar wind erosion, which can result in significant atmospheric loss. Thus,

$$r_{\text{M min}} = R_p. \tag{10}$$

Therefore, the variance of the probability distribution between a probability of 1 and a probability of 0 for this subcriterion is defined as

$$\sigma_{22} = \frac{r_{\text{M early Earth}} - r_{\text{M min}}}{3}, \tag{11}$$

and the probability function is therefore

$$L_{22}(r_M) = \begin{cases} 1 & \text{for } r_M \geqslant 5R_p \\ e^{-\frac{1}{2}\left(\frac{r_M - \mu_{22}}{\sigma_{22}}\right)^2} & \text{for } r_M < 5R_p \end{cases}. \tag{12}$$

*2.3.3. Joint Likelihood for $L_2$ Criteria*

Once the probability functions for a planet retaining its atmosphere through Jeans escape and stellar wind erosion have been defined, it is necessary to analyze the overall probability of atmospheric retention.

To explain our new proposal for combining $L_{21}$ and $L_{22}$, we use the case of early Mars. The epoch called early Mars was between 3700 and 4100 million years ago. There is well-founded evidence that Mars at that time was habitable (B. Sauterey et al. 2022), with an atmosphere much denser than today, and primarily composed of molecular hydrogen (K. Pahlevan et al. 2022); in the southern hemisphere, there were large lakes and rivers, and it is possible that an ocean covered the low-lying plains of the northern hemisphere (G. di Achille & B. M. Hynek 2010). There is also evidence of the existence of a magnetic field that protected the atmosphere from erosion caused by the stellar wind (C. Milbury et al. 2012). The results obtained by applying the habitability index to early Mars have been the following:

$$L_{21} = 0.07, \tag{13}$$
$$L_{22} = 1.00. \tag{14}$$

This result predicts that Mars will have significant thermal escape problems. To determine $L_2$, three options have been considered: the minimum, the maximum, or the average of $L_{21}$ and $L_{22}$. If we use the minimum value, the probability comes out as $L_2 = 0.07$, i.e., the probability that the planet is habitable is almost negligible. However, Mars was habitable for several hundred million years. If we use the maximum of the individual probability functions, $L_2 = 1.00$, the planet would have a very high probability of being habitable, by hiding the problems of thermal escape. If we take the average value, we get a value of $L_2 = 0.53$ that does not hide the fact that the planet has a habitability problem but may be habitable during some time of its life. Therefore, we chose this as the best option. That is,

$$L_2 = \frac{L_{21} + L_{22}}{2}. \quad (15)$$

### 2.4. $L_3$, Supporting Liquid Water on the Surface

The HZ is the circumstellar region where liquid water can remain thermodynamically stable on a planet's surface indefinitely (R. k. Kopparapu et al. 2017). How SEPHI treated the potential habitability of an exoplanet based on this criterion is described in J. M. Rodríguez-Mozos & A. Moya (2017). In SEPHI 2.0, we have introduced a key improvement to the evaluation of this likelihood, the eccentricity of the orbits, which may have a major role in the assessment of habitability.

The normalized luminosity of a star relative to the solar value can be determined as

$$\left(\frac{L_s}{L_\odot}\right) = \left(\frac{R_s}{R_\odot}\right)^2 \left(\frac{T_{\text{eff}}}{T_\odot}\right)^4, \quad (16)$$

where $L_s$ is the luminosity of the star and $L_\odot$ corresponds to the Sun.

On the other hand, the normalized effective flux on a planet, $S_{\text{eff}}$, relative to the flux Earth receives from the Sun, can be determined as

$$S_{\text{eff}} = \frac{L_s}{L_\odot} \frac{1}{a^2}. \quad (17)$$

For elliptical orbits, the normalized mean flux $S'_{\text{eff}}$ received by a planet is given by (R. K. Kopparapu et al. 2013):

$$S'_{\text{eff}} = \frac{S_{\text{eff}}}{\sqrt{1 - e^2}}, \quad (18)$$

where $e$ is the orbital eccentricity. For low eccentricities, the energy flux received is similar to that of a circular orbit. As the eccentricity increases, the mean energy flux per orbit rises, equivalent to a closer circular orbit. To account for the mean impact of this eccentricity, we transform the exoplanet orbit into an equivalent circular orbit radius $a_{\text{eq}}$, defined by

$$a_{\text{eq}} = a(1 - e^2)^{1/4}. \quad (19)$$

Highly eccentric orbits can significantly impact the habitability of exoplanets, affecting their climate over the orbit. Therefore, and following J. M. Rodríguez-Mozos & A. Moya (2017), an exoplanet can be classified into five zones, based on its equivalent circular orbit radius $a_{\text{eq}}$:

1. $a_{\text{eq}} < D_1$: the "hot zone" (water in vapor form);
2. $D_1 \leqslant a_{\text{eq}} < D_2$: the "inner edge" (nonzero probability of water evaporation);
3. $D_2 \leqslant a_{\text{eq}} \leqslant D_3$: the "green zone" (liquid water present);
4. $D_3 < a_{\text{eq}} \leqslant D_4$: the "outer edge" (nonzero probability of $CO_2$ condensation); and
5. $a_{\text{eq}} > D_4$: the "cold zone" (frozen water),

where $D_1$ to $D_4$ are the different limits of these zones in the distance to the hosting star. In the solar system, these points are $(D_1, D_2, D_3, D_4) = (0.51, 0.95, 1.676, 2.4)$ au, corresponding to normalized effective fluxes of $(S_1, S_2, S_3, S_4) = (3.85, 1.107, 0.356, 0.174)$. We use the same procedure as described in J. M. Rodríguez-Mozos & A. Moya (2017) to include the influence of the planetary mass in the definition of these limits and to extrapolate the normalized effective flux of the solar system to other stellar systems. By converting these fluxes into distances from the star, and knowing the equivalent orbital radius $a_{\text{eq}}$ of a planet, we can determine its HZ. The shift of the inner edge of the HZ for synchronously rotating exoplanets around low-mass stars (M and K types) is also considered (R. k. Kopparapu et al. 2016).

From a habitability perspective, exoplanets in the "hot" and "cold" zones have zero probability of hosting liquid water, while those in the "green" zone have a probability of 1. To determine probability functions at the inner edge and the outer edge, we set $\mu_{31} = D_2$ and $\mu_{32} = D_3$, with

$$\sigma_{31} = \frac{D_2 - D_1}{3}, \quad (20)$$

$$\sigma_{32} = \frac{D_4 - D_3}{3}. \quad (21)$$

Thus, the probability function is defined as

$$L_3(a_{\text{eq}}) = \begin{cases} \exp\left(-\frac{1}{2}\left(\frac{a_{\text{eq}} - \mu_{31}}{\sigma_{31}}\right)^2\right) & \text{for } a_{\text{eq}} < D_2 \\ 1 & \text{for } D_2 \leqslant a_{\text{eq}} \leqslant D_3 . \\ \exp\left(-\frac{1}{2}\left(\frac{a_{\text{eq}} - \mu_{32}}{\sigma_{32}}\right)^2\right) & \text{for } a_{\text{eq}} > D_3 \end{cases} \quad (22)$$

### 2.5. SEPHI Compensated

As indicated above, we consider that the values of probability functions cannot be compensated for when determining a habitability index. However, many habitability indexes compensate for the different variables analyzed against each other. In order to compare SEPHI with other habitability indices, an additional index has been created, which we have called "Compensated SEPHI" and which is defined as the harmonic mean of the three probability functions for balancing the influence of the individual subindexes while still prioritizing the most restrictive criterion:

$$\text{SEPHI}_{\text{comp}} = \frac{3}{\frac{1}{L_1} + \frac{1}{L_2} + \frac{1}{L_3}}, \quad (23)$$

where $L_1$, $L_2$, and $L_3$ represent the likelihood functions for being a rocky planet, retaining an atmosphere, and having liquid water, respectively.

## 3. SADE, a Software for Analyzing and Defining Exoplanets

To calculate the SEPHI 2.0 habitability index, a computer program in the Python language has been developed, where all our previous work has been incorporated. The resulting code has been called SADE and its main characteristics are described below.

For a proper assessment of the potential habitability of an exoplanet, SADE performs a comprehensive analysis to characterize each exoplanet and its host star. Specifically, the program determines the type of exoplanet, its internal structure, orbital parameters, HZ analysis, if it is tidally locked, its magnetic properties (assuming an active planetary dynamo), atmospheric escape analysis (thermal and nonthermal), as well as the SEPHI 2.0 habitability index.

**Table 1**
Supplementary Databases Used for Period of Rotation and Age Data of Stars

| Rotation Period | Age |
|---|---|
| T. Reinhold et al. (2023) | Extended Hipparcos Compilation—E. Anderson & C. Francis (2012) |
| E. Díez Alonso et al. (2019) | Geneva Copenhagen Survey—J. Holmberg et al. (2009) |
| R. Angus et al. (2018) | B. Seli et al. (2021) |
| A. Reiners et al. (2018) | L. M. Walkowicz & G. S. Basri (2013) |
| E. R. Newton et al. (2018) | C. C. Worley et al. (2020) |
| L. M. Rebull et al. (2016) | S. Crandall et al. (2020) |
| A. McQuillan et al. (2014) | X. Chen et al. (2020) |
| T. Reinhold et al. (2013) | E. L. Nielsen et al. (2019) |
| M. B. Nielsen et al. (2013) | M. Vioque et al. (2018) |
| N. J. Wright et al. (2011) | K. V. Getman et al. (2014) |
| J. D. Hartman et al. (2010) | J. D. J. do Nascimento et al. (2014) |
| | A. R. Riedel et al. (2014) |
| | I. Ramírez et al. (2013) |
| | M. R. Samal et al. (2012) |
| | G. Bryden et al. (2009) |
| | S. Mathur et al. (2012) |
| | J. H. Debes et al. (2011) |
| | E. L. Shkolnik et al. (2012) |
| | V. A. Marsakov & Y. G. Shevelev (1995) |
| | R. Lachaume et al. (1999) |

### 3.1. SADE Inputs

As input data, we have tried to keep the most basic and common data in the literature for exoplanets and their host stars compatible with assessing the most reliable index possible. To do so, we have taken part of the Plato Data Products as a reference. In particular, SADE needs the following input data, as well as its margins of error:

1. Normalized exoplanet mass.
2. Normalized exoplanet radius.
3. Exoplanet orbital period.
4. Orbital eccentricity.
5. Normalized star mass.
6. Normalized star radius.
7. Star effective temperature.
8. Star rotation period.
9. Star age.

As a demonstrator, we have conducted an online consultation with the NEA database, taking the observational data that NASA currently considers most appropriate. This link to the NASA archive is offered as part of SADE's functionalities.

The star's rotation period is usually not included in exoplanet databases. Therefore, it has been necessary to complement it with other databases that have been included in Table 1.

Likewise, the age of the star system is not usually included in exoplanet databases. Since the NEA includes these data for some stars, when they are absent, they have been supplemented with other databases that have been included in Table 1.

SADE can deal with absent input variables in some particular cases, which are described in detail in the following section, but it cannot analyze exoplanets in cases where there is a simultaneous lack of data relating to the mass and radius or to the orbital period and semimajor axis of the orbit. Nor is it possible to study those stars whose mass, radius, and effective temperature are simultaneously unknown. SADE cannot study the habitability of those stars that are out of range of the limits imposed on the effective temperature in the work of R. K. Kopparapu et al. (2014). This restriction is not a major problem, since stars with spectral groups more massive than F0V have a main-sequence life of less than or equal to 1 Ga, so exoplanets orbiting these stars have a low probability of having enough time for developing life.

### 3.2. SADE Features

The program performs a series of calculations to determine the exoplanet's internal structure, if it is tidally locked, its relation to the HZ, its magnetic properties, and its atmospheric escape velocity. All SADE features are based on previously published works, except for the exoplanet's mass–radius relation, described in Section 4.

In particular, for estimating absent input data, we use the following.

1. For estimating the exoplanet mass:
   – Estimation of the exoplanet mass for radii less than $5.6\,R_\oplus$ from the mass–radius relation included in Section 4 of this paper.
   – Estimation of the exoplanet mass for radii greater than $5.6\,R_\oplus$ from J. Chen & D. Kipping (2017).
2. For estimating the exoplanet radius:
   – Estimation of the exoplanet radius for masses less than $40\,M_\oplus$ from the mass–radius relation included in Section 4 of this paper.
   – Estimation of the exoplanet radius for masses greater than $40\,M_\oplus$ from J. Chen & D. Kipping (2017).
3. For estimating the stellar mass or radius:
   – Estimation of the mass, or radius, of the star from the mass–luminosity relation for intermediate-mass main-sequence stars (O. Y. Malkov 2007).
   – Estimation of the mass, or radius, of the star from the mass–luminosity relation for main-sequence stars (Z. Eker et al. 2015).
   – Estimation of the mass, or radius, of the star from the mass–luminosity relation for main-sequence red dwarfs (G. F. Benedict et al. 2016).
   – Estimation of the mass, or radius, when the luminosity of the star is unknown from the mass–radius relation for main-sequence stars (J. Chen & D. Kipping 2017).
4. For estimating the stellar rotational period:
   – S. A. Barnes (2007) for $0.4 < (B-V) < 0.5$.
   – S. A. Barnes (2010) for $(B-V) < 0.4$ or $(B-V) > 0.9$.
   – E. E. Mamajek & L. A. Hillenbrand (2008) for $0.5 \leqslant (B-V) \leqslant 0.9$.

And for estimating the rest of the exoplanetary system features, we use the following.

1. For the exoplanet's internal structure:
   – Internal structures of dry, water-rich rocky planets from J. M. Rodríguez-Mozos & A. Moya (2022).
   – Compositions of ice and gas-giant planets from J. J. Fortney et al. (2007).
2. For the stellar HZ:

- Determination of the HZ around main-sequence stars from R. K. Kopparapu et al. (2013).
- Influence of the planet's mass on the HZ from R. K. Kopparapu et al. (2014).
- Displacement of the inner edge of the HZ for synchronously rotating planets around low-mass stars (R. k. Kopparapu et al. 2017).
- Influence of the orbital eccentricity on the incident mean effective flux from R. K. Kopparapu et al. (2013).

3. For determining if the exoplanet is tidally locked:
   - Analysis of tidally locking from J. M. Grießmeier et al. (2009).
   - Spin-orbit resonance estimation for tidally locked exoplanets from A. R. Dobrovolskis (2007).

4. For estimating the exoplanet magnetic properties:
   - Estimation of the magnetic regime and moment from J. M. Rodríguez-Mozos & A. Moya (2022).

### 3.3. SADE Results

Once all the necessary data have been determined, SADE performs a Monte Carlo loop with 40,000 executions, allowing the data to vary freely within its margins of error, generating an output set where the following information is displayed:

1. Stellar main characteristics: coordinates, spectral type, mass, radius, effective temperature, luminosity, magnetic regime, age, distance, and rotational period.
2. Exoplanet type and composition: exoplanet name, type of exoplanet (dry rocky, water-rich, etc.), mass, radius, core radius, surface and core–mantle gravity, core, mantle, and mean densities, core mass fraction (CMF), mantle mass fraction (MMF), ice mass fraction (IMF, if applicable), and moment of inertia.
3. Orbit and habitability:
   (a) Orbital parameters: semimajor axis, eccentricity, periastron, apastron, and translation speed.
   (b) Habitability: equivalent semimajor axis, HZ inner radius, HZ outer radius, HZ type, and normalized effective flux.
4. Tidally locked and magnetic properties:
   (a) Tidally locked: tidally locked zone and probability, estimated planet rotation, time synchronization, estimated resonance spin orbit and probability, and planet rotational period.
   (b) Magnetic properties: local Rossby number, magnetic regime, normalized magnetic moment, and surface magnetic field.
5. Exoplanet atmosphere:
   (a) Thermal escape: escape velocity, equilibrium temperature, and loss of components.
   (b) Stellar wind: magnetopause radius, aperture of auroral ring, and type of stellar wind impact.
6. SEPHI 2.0 index: subindexes and the final result.

## 4. Empirical Relations for Estimating Planetary Masses and Radii

More than 90% of confirmed exoplanets have been detected using transit or radial velocity methods. To know the type and composition of an exoplanet, it is essential to have its mass and radius simultaneously. At the moment, only for 18% of confirmed exoplanets, according to NEA, have both variables been determined simultaneously. Therefore, in general, it is necessary to estimate a missing characteristic using the other as a reference. Some of these estimates, such as the one proposed by J. Chen & D. Kipping (2017), have some limitations, including the following:

1. For planets up to $2\,M_\oplus$, only a few data from dry rocky objects in the solar system have been considered when sufficient data from rocky exoplanets are already available that can be considered to determine a more reliable mass–radio relation.

2. From $2\,M_\oplus$, all exoplanet data have been included without differentiating by type when exoplanets of different types, such as dry rocky, water-rich rocky, or ice giants (Neptunian), have different compositions and should maintain differentiated mass–radius ratios.

These limitations led us to develop a new set of empirical relations for estimating exoplanetary masses and radii. Since the training data set (explained in the following sections) lacks observational data in some parts of the parameter space, sometimes the estimations are made by interpolating with a small group of data or even by extrapolating the data set. This can impact the reliability of the estimations. A correct uncertainty estimation can mitigate this problem. This can be addressed via Bayesian inference and by testing how this model can generalize the estimations, but these improvements are out of the scope of this work and will be part of the future improvements of SADE.

### 4.1. Mass–Radius Relationship for Exoplanets Up to $40\,M_\oplus$

To establish a mass–radius relationship for exoplanets up to $40\,M_\oplus$, we assume that gas giants are unlikely to be present within this mass range. Hence, only dry rocky planets, water-rich rocky planets, and Neptune-like planets are considered.

The data provided by NEA have been taken from all the exoplanets confirmed prior to the date of this study, around 5500 planets, where the mass and radius are simultaneously known. A Monte Carlo loop is performed on this data set, allowing the variables to vary randomly within their observational uncertainties. Each exoplanet is then placed in our mass–radius diagram (J. M. Rodríguez-Mozos & A. Moya 2022) to determine the most probable planet type, mass, radius, and internal composition, without relying on any estimate. Additionally, we include data from the solar system planets that belong to the considered types.

A series of filters is applied to this set of planets, to ensure a high-quality mass–radius relationship. Thus, the following cases are excluded:

1. Gas giants.
2. Exoplanets with less than a 55% probability of belonging to one of the analyzed types.
3. Those exoplanets whose error in determining the radius is greater than 15% or the error in determining the mass is greater than 40%.
4. Rocky planets with a CMF exceeding the collisional stripping curve, as such planets may have undergone dramatic events, making their data nonrepresentative.

After applying these conditions, a final set of 216 planets is obtained (see Table 2, available in the online edition). These planets are plotted on a logarithmic mass–radius diagram (Figure 1). Based on this figure, the following conclusions can be drawn:

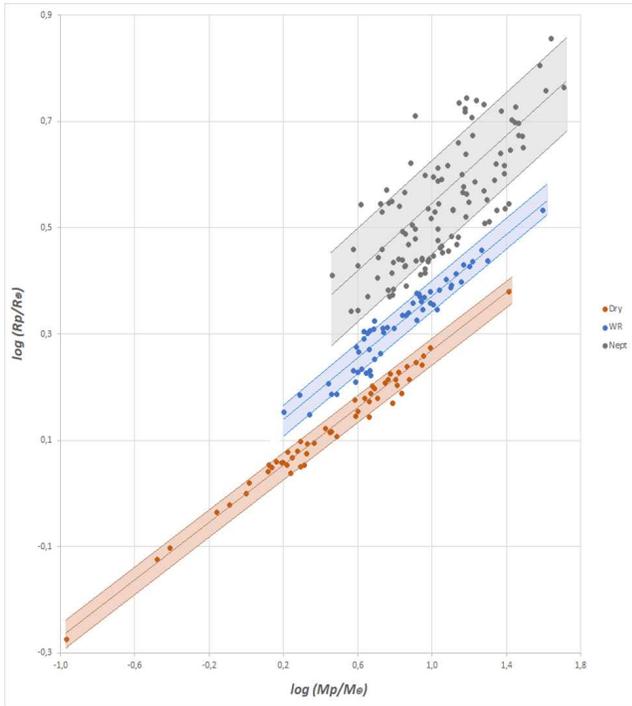

**Figure 1.** Mass–radius diagram featuring three exoplanet populations. The brown points represent the dry rocky exoplanet population, while the blue and gray points correspond to water-rich rocky planets and Neptune-like planets, respectively. The straight lines denote the power-law relation derived for each population (mean ± standard deviation), indicating a 68% probability that an exoplanet of a given type lies within the shaded region.

**Table 2**
Short Excerpt of the Data—the Full Table is Published Online

| Planet | Type | Prob Type | Mass | Mass Error (%) | Radius | Radius Error (%) |
|---|---|---|---|---|---|---|
| Mars | Dry | 1.0 | 0.107 | 0.1 | 0.532 | 0.1 |
| TRAPPIST-1 h | Dry | 0.73 | 0.33 | 6.1 | 0.752 | 1.9 |
| TRAPPIST-1 d | Dry | 0.96 | 0.388 | 3.1 | 0.788 | 1.3 |
| Venus | Dry | 1.0 | 0.815 | 0.1 | 0.95 | 0.1 |
| Earth | Dry | 1.0 | 1.0 | 0.0 | 1.0 | 0.0 |
| … | … | … | … | … | … | … |

**Note.** A portion is shown here for guidance regarding its form and content. The entire table is available in machine-readable form in the online journal.

(This table is available in its entirety in machine-readable form in the online article.)

1. Three distinct exoplanet populations coexist in this mass range: dry rocky planets, water-rich rocky planets, and Neptune-like planets.
2. Determining to which population an exoplanet belongs requires simultaneous knowledge of both its mass and radius.
3. The mean masses of each population differ significantly. Dry rocky planets have a mean mass of $4.1 \pm 3.9\,M_\oplus$, water-rich rocky planets have a mean mass of $8.2 \pm 6.1\,M_\oplus$, and Neptune-like planets have a mean mass of $13.9 \pm 7.2\,M_\oplus$.

### 4.2. Determination of Exoplanet Radius

Using the data for the 216 exoplanets, we perform an approximation for each of the exoplanet populations to determine the radius based on mass using a power law of the form:

$$\left(\frac{R_p}{R_\oplus}\right) = C\left(\frac{M_p}{M_\oplus}\right)^S, \quad (24)$$

where $M_p$ and $R_p$ are the mass and radius of the exoplanet, $M_\oplus$ and $R_\oplus$ are Earth's mass and radius, and $C$ and $S$ are the parameters that describe the power law. The best-fit values for the three populations are presented in Table 3.

The power law derived for dry rocky planets is consistent with the one determined by L. Zeng et al. (2019), despite their sample being smaller and slightly more massive than ours, and it closely matches the relation found by J. Chen & D. Kipping (2017).

Considering the mass distributions listed in Table 3, we can statistically estimate the most likely population to which an exoplanet belongs. Figure 2 shows the probability of an exoplanet belonging to each of the three populations as a function of mass, assuming that the masses follow normal distributions defined by the parameters indicated in Table 3.

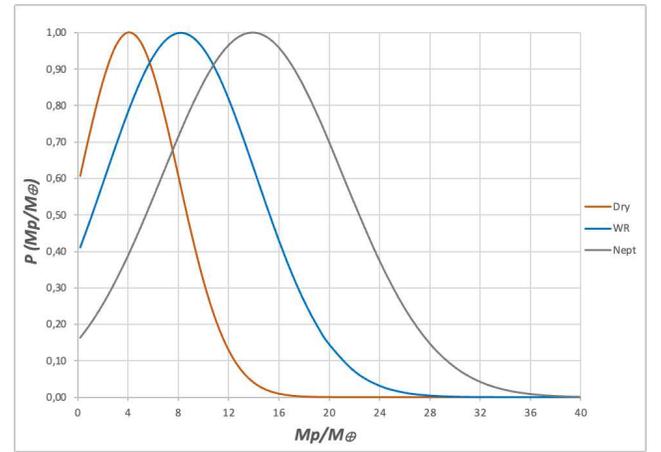

**Figure 2.** Probability of an exoplanet belonging to each population as a function of mass, based on the mass distributions from Table 3.

**Table 3**
Power-law Parameters for Each Exoplanet Population

| Type of Planet | C | S | Error (%) | Mass Distribution $N(\mu \pm \sigma)$ |
|---|---|---|---|---|
| Dry Rocky Planets | 1.00 | 0.27 | 4 | $4.1 \pm 3.9$ |
| Water-rich Rocky Planets | 1.20 | 0.29 | 5 | $8.2 \pm 6.1$ |
| Neptunian Worlds | 1.68 | 0.32 | 19 | $13.9 \pm 7.2$ |

### 4.3. Determination of Exoplanet Mass

A similar analysis to the one in the previous section was conducted for each exoplanet population to determine the mass based on radius, using a power law of the form

$$\left(\frac{M_p}{M_\oplus}\right) = D\left(\frac{R_p}{R_\oplus}\right)^T, \quad (25)$$

where $D$ and $T$ are the parameters that define the power law; the best-fit values are presented in Table 4.

**Table 4**
Power-law Parameters for Each Exoplanet Population

| Type of Planet | D | T | Error (%) | Radius Distribution $N(\mu \pm \sigma)$ |
|---|---|---|---|---|
| Dry Rocky Planets | 1.00 | 3.70 | 14 | $1.3 \pm 0.3$ |
| Water-rich Rocky Planets | 0.49 | 3.42 | 20 | $2.1 \pm 0.4$ |
| Neptunian Worlds | 1.29 | 1.79 | 44 | $3.6 \pm 0.7$ |

Considering the radius distributions listed in Table 4, it is possible to estimate the most likely population to which an exoplanet belongs based on its radius. Figure 3 shows the probability of an exoplanet belonging to each population as a function of radius, assuming normally distributed radii based on the parameters in Table 4.

### 4.4. Probability Change Points

We define probability change points (PCPs) as those points where the dominant probability of belonging to one exoplanet population shifts to another. Each PCP is defined by two coordinates: mass and radius. In Figure 3, we observe that for small exoplanet radii, the predominant probability is that of being a dry rocky planet. However, there is a cutoff point between the probability curves, labeled PCP1, beyond which it becomes more likely for the exoplanet to be water-rich. Similarly, this occurs for the point of probability change from water-rich to Neptunian exoplanets, which has been called PCP2.

To determine the values of these change points, we utilize the probability distributions obtained for the exoplanet radius in Table 4. This choice is justified by the fact that the average observational error in the radius determination for the 216 exoplanets used in this study is 4.4%, while the average observational error in the mass determination is 16.1%. Thus, PCP1 is determined by calculating the intersection of the brown and blue curves in Figure 3, with the associated mass obtained from the power law for dry rocky planets. Consequently, PCP1 is defined by the following coordinates:

$$\text{PCP1} = (6.3, 1.64), \qquad (26)$$

where 6.3 is the mass and 1.64 is the radius at the PCP. An analogous procedure can be followed to determine PCP2, by calculating the intersection of the blue and gray curves in Figure 3, leaving PCP2 to be defined by the following coordinates:

$$\text{PCP2} = (13.7, 2.65). \qquad (27)$$

## 5. Application to the Solar System and Other Objects of Interest

The SADE program has been applied with the aim of knowing the habitability index of SEPHI 2.0 of all the planets in the solar system and other objects that are considered very interesting, such as the Galilean moons of Jupiter, the exoplanets belonging to the TRAPPIST-1 star system, and the closest exoplanet to Earth, Proxima Centauri b.

The results obtained have been included in Table 5. In this table, it can be seen that all the inner planets of the solar system are dry rocky planets. SEPHI 2.0 reveals that Mercury is not habitable, because it is outside the HZ of the Sun, while Venus has a low SEPHI value (0.277), because there is a high probability that the water is in the form of vapor. Earth has a SEPHI 2.0 value of 0.945, which may initially be shocking, because we are used to thinking of the Earth as a model of habitability; however, our planet has a small problem, since it does not have enough mass to retain the hydrogen and helium molecules that leave the atmosphere through thermal escape mechanisms. For Mars, the SEPHI 2.0 and $\text{SEPHI}_{\text{comp}}$ indexes indicate that it may be habitable. As we have seen above, it was very likely habitable for a time; however, the problem of thermal escape is much greater than on Earth, because molecules of hydrogen, helium, ammonia, methane, and even water vapor can leave the atmosphere.

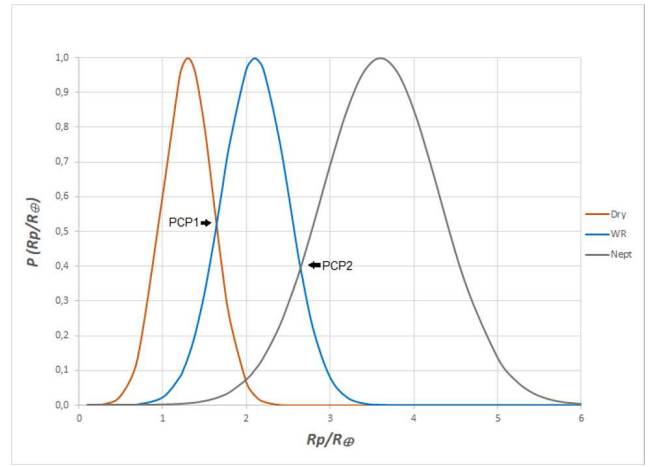

**Figure 3.** Probability of an exoplanet belonging to each population as a function of radius, based on the radius distributions from Table 4.

As for the outer planets of the solar system—Jupiter, Saturn, Uranus, and Neptune—SEPHI 2.0 reveals that they are not habitable for two reasons: they are not rocky planets and they are not found in the HZ of the Sun. It has also been considered interesting to calculate the habitability index for the moons of Jupiter discovered by Galileo—Io, Europa, Ganymede and Callisto. Evidently, the four moons are outside the HZ and are therefore not habitable, although if they had had different orbits, they could have been habitable. Specifically, Ganymede, which is similar in size to Mercury, with an orbit within the HZ, could have been a habitable ocean planet for a time during its lifetime, although the lack of sufficient mass to retain the atmosphere would make it uninhabitable over the long term.

Once it was verified that SEPHI 2.0 adequately describes the reality of the solar system, the habitability indices of the TRAPPIST-1 and Proxima Centauri stellar systems were determined. These exoplanets are located close to Earth and have been studied in depth, so their observational data are quite reliable. The TRAPPIST-1 planets are very interesting, because they are dry rocky planets similar in size to Earth. SEPHI 2.0 prioritizes planets f and g that could be habitable; however, SADE describes some problems that can endanger their habitability. On the one hand, the high activity of the star is a problem in itself. This star rotates with a very fast period (3.3 days), generating a powerful and dangerous stellar wind to maintain the planets' atmospheres. On the other hand, in low-mass red dwarfs, such as TRAPPIST-1, the planets located in the HZ are tidally locked with their star, so that the planetary magnetic field, if the planetary dynamo is still active, will be weak and probably unable to cope with the erosion caused by

**Table 5**
SEPHI 2.0 and SEPHI$_{comp}$ Calculations for Solar System Planets and Moons and Exoplanets of Interest

| Object | Mass | Radius | Type | $L_1$ | $L_2$ | $L_3$ | SEPHI 2.0 | SEPHI$_{comp}$ |
|---|---|---|---|---|---|---|---|---|
| Mercury | 0.055 | 0.38 | D | 1.00 | 0.02 | 0.00 | 0.000 | 0.000 |
| Venus | 0.815 | 0.95 | D | 1.00 | 0.80 | 0.28 | 0.277 | 0.515 |
| Earth | 1.00 | 1.00 | D | 1.00 | 0.95 | 1.00 | 0.945 | 0.981 |
| Mars | 0.107 | 0.53 | D | 1.00 | 0.53 | 1.00 | 0.531 | 0.772 |
| Jupiter | 317.9 | 10.97 | G | 0.00 | 1,00 | 0.00 | 0.000 | 0.000 |
| Saturn | 95.20 | 9.14 | G | 0.00 | 1.00 | 0.00 | 0.000 | 0.000 |
| Uranus | 14.54 | 3.98 | N | 0.00 | 1.00 | 0.00 | 0.000 | 0.000 |
| Neptune | 17.14 | 3.86 | N | 0.00 | 1.00 | 0.00 | 0.000 | 0.000 |
| Io | 0.015 | 0.29 | D | 1.00 | 0.51 | 0.00 | 0.000 | 0.000 |
| Europa | 0.008 | 0.25 | WR | 0.87 | 0.51 | 0.00 | 0.000 | 0.000 |
| Ganymede | 0.025 | 0.41 | WR | 0.82 | 0.51 | 0.00 | 0.000 | 0.000 |
| Callisto | 0.018 | 0.38 | WR | 0.68 | 0.51 | 0.00 | 0.000 | 0.000 |
| TRAPPIST-1 b | 1.37 | 1.12 | D | 1.00 | 0.38 | 0.00 | 0.000 | 0.000 |
| TRAPPIST-1 c | 1.31 | 1.10 | D | 1.00 | 0.44 | 0.00 | 0.000 | 0.000 |
| TRAPPIST-1 d | 0.388 | 0.79 | D | 1.00 | 0.13 | 0.02 | 0.023 | 0.058 |
| TRAPPIST-1 e | 0.692 | 0.92 | D | 1.00 | 0.36 | 1.00 | 0.362 | 0.628 |
| TRAPPIST-1 f | 1.04 | 1.05 | D | 1.00 | 0.50 | 1.00 | 0.499 | 0.750 |
| TRAPPIST-1 g | 1.32 | 1.13 | D | 1.00 | 0.50 | 1.00 | 0.500 | 0.750 |
| TRAPPIST-1 h | 0.330 | 0.75 | D | 1.00 | 0.28 | 0.54 | 0.282 | 0.467 |
| Proxima Cen b | 1.07 | 1.00 | D | 1.00 | 0.50 | 1.00 | 0.504 | 0.752 |

**Note.** The mass and radius are, respectively, the mass and radius of the object normalized to the terrestrial value. The type is the type of object, with "D" being dry rocky objects, "WR" being water-rich rocky objects, and "N" being Neptunian or ice-giant objects. $L_1$, $L_2$, and $L_3$ are, respectively, the probability functions of being a rocky object, maintaining the atmosphere, and having liquid water on the surface.

the stellar wind. For this reason, $L_2$ function values close to 0.5 reflect the difficulty in maintaining the atmosphere due to nonthermal escape. A similar case occurs with Proxima Centauri b: it is also a rocky planet located in the HZ of its star that also presents problems with nonthermal escape. The SEPHI 2.0 value for this exoplanet is slightly better, due to a higher magnetic field that will better protect the atmosphere from the stellar wind.

## 6. Application to Confirmed Exoplanets

SADE has been applied to all confirmed exoplanets to date (2024 August), determining the SEPHI 2.0 habitability index for each exoplanet based on data from NEA. It is worth noting that for some exoplanets, only upper limits on their mass or radius are known. For example, the mass of Kepler-62 f is known to be less than 35 $M_\oplus$, but applying our mass-to-radius ratio using its radius (1.41 $R_\oplus$), we estimate a more likely mass of around 3.6 $M_\oplus$, which is well below that upper limit. In such cases, the introduction of the upper limit to the Monte Carlo loop could lead to errors in determining the type of planet and its most likely composition. Therefore, when only an upper limit is available for mass or radius, this upper limit is not taken into account, and our mass–radius relations are directly applied to estimate the mass from the radius or vice versa.

The results for the 26 exoplanets with the highest habitability indices are included in Table 6. The SEPHI 2.0 habitability index should be viewed as a probability, meaning any index value of 0.5 or greater is worth considering for more detailed follow-up analysis.

From these results, it can be concluded that the new SEPHI 2.0 habitability index is very strict in the evaluation of planetary habitability. Among the more than 5500 exoplanets confirmed to date, only two planets, Kepler-62 f and GJ 514 b, have habitability indices close to 1. In addition, the SEPHI 2.0 indices for Kepler-441 b (0.874) and Kepler-452 b (0.864) reflect the uncertainties regarding their classification as dry or water-rich rocky planets. This uncertainty could be resolved once their masses are determined, eliminating the need to estimate them.

It should be noted that the exoplanet Kepler-62 f, with the observational data available at this time, meets all the conditions imposed by SEPHI 2.0—that is, it is a rocky planet, which can maintain its atmosphere against stellar wind erosion and thermal escape, and is located in the HZ of its star.

An exoplanet will undoubtedly be more or less habitable regardless of its distance from the solar system. However, Table 6 has also included the distance at which the exoplanet is located, in order to prioritize habitability studies for those exoplanets that are closer and therefore more accessible observationally for atmospheric characterization. In this way, observations with the current JWST mission or with the future Habitable Worlds Observatory can be prioritized.

## 7. Comparison of SEPHI 2.0 with Other Indices

### 7.1. Comparison with ESI

ESI (D. Schulze-Makuch et al. 2011), in its simplified formulation, is defined as

$$\mathrm{ESI} = D_1 . D_2 . D_3 . D_4, \tag{28}$$

where

$$D_i = \left(1 - \left|\frac{x_i - 1}{x_i + 1}\right|\right)^{\frac{w_i}{4}}. \tag{29}$$

$D_i$ is, therefore, the deviation from the unit value of the normalized variable $x_i$ of the exoplanet, and $w_i$ is the weight assigned to this variable within the index. ESI calculates the differences of four variables—radius, density, escape velocity, and surface temperature—of an exoplanet in relation to terrestrial values.

Table 6
Results for Exoplanets with Better Habitability Index via SEPHI 2.0

| Exoplanet | Type[a] | Mp[b] | Rp[c] | CMF | MMF | IMF | $a_{eq}$[d] | D2 | D3 | Zone[e] | $S_{eff}$ | Rot[f] | Reg.[g] | MM[h] | SEPHI$_{2.0}$ | SEPHI$_{comp}$ | Distance |
|---|---|---|---|---|---|---|---|---|---|---|---|---|---|---|---|---|---|
| Kepler-62 f | Dry | 3.6 ± 0.3 | 1.41 ± 0.04 | 0.34 | 0.66 | 0 | 0.718 | 0.452 | 0.840 | G | 0.42 | F | D | 4.6 | 1.000 | 1.000 | 301 |
| GJ 514 b | Dry | 5.3 ± 0.5 | 1.54 ± 0.08 | 0.39 | 0.61 | 0 | 0.389 | 0.208 | 0.412 | G | 0.29 | F | D | 7.5 | 0.994 | 0.998 | 7.6 |
| Kepler-441 b | Dry | 6.1 ± 0.5 | 1.57 ± 0.10 | 0.45 | 0.55 | 0 | 0.566 | 0.302 | 0.583 | G | 0.30 | F | D | 9.6 | 0.874 | 0.949 | 268 |
| Kepler-452 b | WR | 2.7 ± 0.3 | 1.63 ± 0.12 | 0.08 | 0.46 | 0.46 | 1.049 | 1.061 | 1.901 | IE | 1.17 | F | D | 0.9 | 0.865 | 0.949 | 552 |
| HD 40307 g | WR | 7.3 ± 1.4 | 2.11 ± 0.12 | 0.08 | 0.47 | 0.45 | 0.592 | 0.495 | 0.930 | G | 0.76 | F | DR | 2.6 | 0.831 | 0.927 | 12.9 |
| Kepler-1544 b | WR | 3.6 ± 0.4 | 1.79 ± 0.05 | 0.07 | 0.41 | 0.52 | 0.556 | 0.517 | 0.962 | G | 0.91 | F | D | 1.0 | 0.725 | 0.837 | 335 |
| HD 216520 c | WR | 9.5 ± 0.9 | 2.29 ± 0.13 | 0.07 | 0.41 | 0.52 | 0.525 | 0.565 | 1.054 | IE | 1.27 | F | DR | 2.9 | 0.709 | 0.850 | 19.6 |
| Kepler-1638 b | WR | 4.6 ± 0.5 | 1.89 ± 0.14 | 0.08 | 0.45 | 0.47 | 0.786 | 0.896 | 1.635 | IE | 1.53 | F | D | 1.4 | 0.679 | 0.776 | 1526 |
| HD 31527 d | WR | 11.9 ± 1.0 | 2.45 ± 0.14 | 0.06 | 0.38 | 0.56 | 0.791 | 0.801 | 1.459 | IE | 1.24 | F | D | 3.0 | 0.653 | 0.834 | 38.4 |
| Kepler-22 b | WR | 6.2 ± 0.7 | 2.10 ± 0.07 | 0.06 | 0.35 | 0.59 | 0.798 | 0.756 | 1.389 | G | 1.04 | F | D | 1.5 | 0.625 | 0.825 | 195 |
| Kepler-1606 b | WR | 6.4 ± 0.7 | 2.10 ± 0.11 | 0.06 | 0.37 | 0.57 | 0.638 | 0.721 | 1.331 | IE | 1.46 | F | D | 1.6 | 0.621 | 0.730 | 831 |
| Kepler-174 d | WR | 7.2 ± 0.8 | 2.19 ± 0.07 | 0.06 | 0.34 | 0.61 | 0.692 | 0.426 | 0.802 | G | 0.41 | F | D | 1.6 | 0.601 | 0.771 | 385 |
| Kepler-1701 b | WR | 7.6 ± 0.8 | 2.20 ± 0.12 | 0.06 | 0.36 | 0.58 | 0.571 | 0.633 | 1.180 | IE | 1.37 | F | D | 1.9 | 0.589 | 0.735 | 584 |
| Kepler-1090 b | WR | 8.2 ± 0.9 | 2.26 ± 0.11 | 0.06 | 0.34 | 0.60 | 0.634 | 0.663 | 1.228 | IE | 1.90 | F | D | 1.9 | 0.575 | 0.739 | 859 |
| Kepler-443 b | WR | 8.8 ± 1.0 | 2.30 ± 0.11 | 0.06 | 0.34 | 0.60 | 0.557 | 0.463 | 0.878 | G | 0.73 | F | D | 2.0 | 0.560 | 0.723 | 802 |
| Kepler-69 c | WR | 3.4 ± 0.4 | 1.74 ± 0.14 | 0.08 | 0.46 | 0.46 | 0.707 | 0.857 | 1.550 | IE | 1.69 | F | D | 1.1 | 0.558 | 0.738 | 731 |
| GJ 180 c | Dry | 7.4 ± 1.7 | 1.62 ± 0.09 | 0.51 | 0.49 | 0 | 0.123 | 0.126 | 0.256 | IE | 1.06 | S | M | 0.1 | 0.530 | 0.661 | 11.9 |
| GJ 682 c | Dry | 5.9 ± 1.3 | 1.53 ± 0.09 | 0.52 | 0.48 | 0 | 0.085 | 0.067 | 0.138 | G | 0.63 | S | M | 0.11 | 0.526 | 0.754 | 5.0 |
| HD 20794 e | Dry | 4.9 ± 0.5 | 1.51 ± 0.08 | 0.40 | 0.60 | 0 | 0.498 | 0.564 | 1.601 | IE | 1.37 | SS | M | 0.02 | 0.525 | 0.709 | 6.0 |
| GJ 667 C f | Dry | 3.1 ± 0.6 | 1.30 ± 0.07 | 0.50 | 0.50 | 0 | 0.155 | 0.083 | 0.165 | G | 0.28 | S | M | 0.03 | 0.518 | 0.763 | 7.2 |
| GJ 433 d | Dry | 5.3 ± 0.5 | 1.54 ± 0.08 | 0.39 | 0.61 | 0 | 0.167 | 0.165 | 0.336 | G | 0.98 | S | M | 0.04 | 0.518 | 0.729 | 9.1 |
| GJ 367 d | Dry | 6.1 ± 0.3 | 1.60 ± 0.08 | 0.37 | 0.63 | 0 | 0.159 | 0.171 | 0.343 | IE | 1.16 | S | M | 0.04 | 0.517 | 0.629 | 9.4 |
| Kepler-442 b | Dry | 2.7 ± 0.2 | 1.28 ± 0.07 | 0.43 | 0.57 | 0 | 0.386 | 0.349 | 0.657 | G | 0.82 | S | M | 0.01 | 0.516 | 0.762 | 366 |
| tau Cet e | Dry | 4.1 ± 0.4 | 1.44 ± 0.08 | 0.40 | 0.60 | 0 | 0.535 | 0.520 | 0.980 | G | 0.99 | S | M | 0.01 | 0.516 | 0.760 | 3.6 |
| GJ 667 C c | Dry | 4.1 ± 0.7 | 1.43 ± 0.08 | 0.44 | 0.56 | 0 | 0.125 | 0.082 | 0.165 | G | 0.43 | S | M | 0.04 | 0.516 | 0.762 | 7.2 |
| Kepler-1652 b | Dry | 5.7 ± 0.5 | 1.57 ± 0.09 | 0.40 | 0.60 | 0 | 0.162 | 0.150 | 0.300 | G | 0.86 | S | M | 0.04 | 0.516 | 0.739 | ⋯ |

**Notes.**
[a] Dry = dry rocky planet; WR = water-rich rocky planet.
[b] Mp = normalized planet mass.
[c] Rp = normalized planet radius.
[d] $a_{eq}$ = equivalent semimajor axis.
[e] G = green; IE = inner edge; OE = outer edge.
[f] F = free; S = synchronous; SS = super-synchronous.
[g] D = dipolar; DR = dipolar reversing; M = multipolar.
[h] MM = normalized magnetic moment.
[i] Distance in parsecs.

When ESI is applied to the planets of the solar system, the result obtained for Mercury is frankly disappointing (0.6), since it is very similar to Mars (0.7), with Mars being a planet located in the HZ that was probably habitable during a stage of its history.

On the other hand, ESI has other added problems:

1. ESI is not a habitability index in strict terms; it only represents the difference of four variables with terrestrial values. An exoplanet larger or denser than Earth could be even more habitable than our planet, and ESI would only pick up the differences with terrestrial values.

2. The weights attributed to each variable, rather than to equalize them, are used to prioritize them in the global index, finally resulting in an index that is extremely sensitive to surface temperature.

### 7.2. Comparison with PHI

PHI (D. Schulze-Makuch et al. 2011) is defined as

$$\mathrm{PHI} = (S * E * C * L)^{\frac{1}{4}}, \qquad (30)$$

with S, E, C, and L being the conditions of habitability. Condition S is to present a stable substrate, where it is valued that the exoplanet has a solid surface, an atmosphere, and a magnetosphere. Condition E is to have energy sources and the existence of starlight, redox reactions, and tidal heating. Condition C is to have an appropriate chemistry and the presence of organic molecules, nitrogen, sulfur, and phosphorus. And condition L is to have a solvent liquid and its presence in the atmosphere, on the surface and below the surface.

PHI applied to the solar system produces quite reasonable results: Mercury (0), Venus (0.39), Earth (0.97), and Mars (0.56). PHI is a very theoretical index that, in general, cannot be applied to exoplanets, since there are many unknown aspects, such as the existence of redox reactions, the presence of organic molecules, sulfur, or phosphorus atoms, and solvent liquid on the surface or below it, etc. However, SEPHI 2.0 is a habitability index designed to be applied online to any exoplanet, based on the available observational data.

### 7.3. Comparison with the Habitability Index for Transiting Exoplanets

The Habitability Index for Transiting Exoplanets (HITE; R. Barnes et al. 2015) is a serious attempt to analyze the habitability of exoplanets discovered by photometric transit.

According to HITE, for a planet to be habitable, it must simultaneously be: (i) a rocky planet; and (ii) emit a flow of energy within the limits of the HZ. The input data for applying HITE are the planet's orbital period, the depth and duration of the transit, and the impact parameter, when known. As for the data from the star, HITE needs the radius, the effective temperature, and the logarithm of the surface gravity. On the other hand, the degeneracy between the eccentricity of the orbit (e) and the albedo (A) of the planet is solved in a very elegant way, with HITE representing the sets of combinations (A, e) that make the planet habitable against the total combinations evaluated, and therefore the result is given in terms of probability.

The HITE values, without taking into account the eccentricity of the orbit, for Venus, Earth, and Mars are, respectively, 0.300, 0.829, and 0.422. These results are somewhat lower than those obtained by SEPHI 2.0 for Earth (0.945) and Mars (0.531).

The results obtained by HITE and SEPHI for confirmed exoplanets have also been compared. In many cases, the results of both indices are similar and the conclusions obtained are the same—for example, for Kepler-442 b (HITE 0.838/SEPHI$_{\mathrm{comp}}$ 0.765), Kepler-1652 b (HITE 0.913/ SEPHI$_{\mathrm{comp}}$ 0.717), and Kepler-62 f (HITE 0.650/ SEPHI$_{\mathrm{comp}}$ 1.000). In these exoplanets, the differences in the values of the indexes are basically due to the atmosphere, an issue studied by SEPHI but not HITE. In other cases, such as Kepler-296 e (HITE 0.850/ SEPHI$_{\mathrm{comp}}$ 0.407), the difference is due to the value of the eccentricity ($e = 0.164$), which makes the normalized effective flux received by the exoplanet (Seff = 1.50) significantly higher than that considered by HITE (Seff = 1.08).

The most important differences found between HITE and SEPHI 2.0 are the following:

1. HITE is only applicable to exoplanets discovered by transit, while SEPHI 2.0 is applicable to any confirmed exoplanet, regardless of the detection method.

2. The mass-to-radius relation used by HITE only differentiates between planets with radii greater or smaller than Earth. The mass–radius ratio used by SEPHI 2.0, and described in this paper, varies according to the type of planet and is considered much more complete.

3. To determine the probability that a planet is rocky, HITE only considers the variable radius of the planet. Actually, the right way to determine if a planet is rocky is to enter its mass and radius in a mass–radius diagram, which is the method followed by SEPHI 2.0.

## 8. Conclusions

In this paper, an improved version of the SEPHI 2.0 habitability index and a complete program for a comprehensive characterization of exoplanets are presented. These tools allow for an automated assessment of the characteristics of each exoplanet as well as the habitability problems that may exist.

In this way, the main achievements of this work can be listed as follows:

1. A new version of the SEPHI 2.0 habitability index is proposed, where the composition of the exoplanet, the ability to retain the atmosphere by studying thermal escapes, and the erosion produced by the stellar wind are analyzed in greater depth, as well as the possibility that the exoplanet has liquid water on its surface. The new index is very demanding compared to other habitability indexes, in particular the previous version of SEPHI. In this version, we have modified the following features:

—The way we combine the different subcriteria for obtaining the final index. Now we select the most restrictive index as the final result.

—We have changed the way we evaluate the probability of the exoplanet retaining an atmosphere, now combining information from thermal (Jeans escape) and nonthermal (stellar winds and protecting magnetic field) processes.

—The orbital eccentricity is now included, for determining the potential of the exoplanet to have liquid water.

In this context, only one exoplanet, Kepler-62 f, presents the unit value of the habitability index. We have also presented an alternative subindex combination, called SEPHI$_{\mathrm{comp}}$, providing another approximation to the problem, and slightly different results, without changing the main conclusions. SEPHI$_{comp}$ is also suitable for comparing with other indexes in the literature.

2. A new mass–radius relationship obtained from the observational data of exoplanets has been presented. The training data set is complemented by the data of the solar system planets. It has been found that three differentiated populations of exoplanets coexist in the analyzed mass range: dry rocky planets, water-rich rocky planets, and ice giants or Neptunian planets. A mass-to-radius ratio has been determined for each one of the three populations, and the problem has been analyzed from a statistical point of view, to determine where an exoplanet is most likely to be a dry rocky planet, a water-rich rocky planet, or a Neptunian-like planet. This led us to the definition of the PCPs—the points in the mass–radius plane where the exoplanet changes its population from a statistical point of view.

3. The SADE program has been presented, which can be accessed through the following website: https://sadeweb.azurewebsites.net/. This allows exoplanets to be analyzed automatically, as well as to determine their habitability index from the available observational data. We have applied it to all the exoplanets in the NEA and the solar system. SADE and SEPHI 2.0 can help to prioritize additional observations for the most interesting exoplanets and, at the same time, to discard those that may have significant habitability problems.

## Acknowledgments

The authors thank the anonymous referee for the very valuable and constructive comments that helped to improve the initial version of this work. A.M. acknowledges funding support from grant PID2019-107061GB-C65, funded by MCIN/AEI/10.13039/501100011033, and from Generalitat Valenciana in the frame of the GenT Project CIDEGENT/2020/036.

## ORCID iDs

A. Moya  https://orcid.org/0000-0003-1665-5389